\documentclass{iopart}
\usepackage{iopams,bm,mathrsfs}
\usepackage[mathcal]{euscript}
\usepackage{temp}
\usepackage{hyperref}
\usepackage[numbers,sort&compress]{natbib}
\renewcommand{\vec}[1]{\bm{#1}}
\renewcommand{\imath}{\rmi}
\date{September 27, 2006}

\begin{document}

\title[Field momentum and gyroscopic dynamics\dots ]{Field momentum
and gyroscopic dynamics of classical systems with topological defects}

\author{Denis D. Sheka}

\ead{Denis\_Sheka@univ.kiev.ua}

\address{National Taras Shevchenko University of Kiev, 03127 Kiev, Ukraine}

\begin{abstract}
The standard relation between the field momentum and the force is generalized
for the system with a field singularity: in addition to the regular force,
there appears the singular one. This approach is applied to the description of
the gyroscopic dynamics of the classical field with topological defects. The
collective--variable Lagrangian description is considered for gyroscopical
systems taking into account singularities. Using this method we describe the
dynamics of two--dimensional magnetic solitons. We establish a relation
between the gyroscopic force and the singular one. An effective Lagrangian
description is discussed for the magnetic soliton dynamics.
\end{abstract}

%

%
%
\submitto{JPA}

\maketitle


\section{Introduction}
\label{sec:introduction} %

An important role in the modern physics of condense matter and in field
theories is connected with the study of topological defects. Common examples
are specific distributions of the order parameter as dislocations,
disclinations, vortices, monopoles, hedgehogs, boojums, etc. At present the
topological classification of defects is almost done, but the dynamical theory
of defects is far from completeness. In the continuum approach, defects are
described by essentially nonlinear solutions of field equations like
topological solitons. The classical field theory in its standard form is
suitable only for analysis of fields, which can be described by regular
functions, while the soliton profile can be singular. This leads to
ambiguities in the energy--momentum tensor problem: the linear momentum is
either not well defined or is not conserved. Typical examples for this
long--standing paradox in the condense matter theory are the magnetism, where
there is no well--defined energy--momentum tensor; the canonical definition
for the field momentum fails for the magnetic bubbles \cite{Slonczewski79};
this canonical momentum is not invariant under spin rotations
\cite{Haldane86}. The part of the problem, which is connected with the absence
of the momentum invariance under gauge transformation, can be explained on the
microscopic quantum level as a result of momentum exchange with microscopic
degrees of freedom \cite{Volovik87}. In this case, it is possible to treat the
problem by introducing the nonlocal Novikov--Wess--Zumino term in the action
\cite{Volovik87,Volovik03}.

A new discussion of the momentum problem appeared in the last decade due to
the study of dynamics of topological solitons in low--dimensional magnetism.
In particular, usage of canonical momentum for the construction of the
effective equation of motion for magnetic vortices leads to contradictions
between different approaches
\cite{Nikiforov83,Papanicolaou91,Wysin94a,Mertens00}. Let us note that the
problem was solved by \citet{Papanicolaou91} for the special case of localized
magnetic solitons (often named `skyrmions'), where the nonstandard form of
field momentum was constructed as a moment of vorticity; however, such
approach is not universal.

In this paper, we show how to avoid the problem in terms of standardly defined
field momentum. We construct the equation of motion, involving the force and
the momentum, which is suitable for the description of singular objects like
topological defects. By generalizing an equation for the energy--momentum
flux, we calculate the relation between the time derivative of the momentum
and the force acting the system (Newtonian--like equation). We prove in
section \ref{sec:balance} that in addition to the regular force there appears
the singular one, which exists in the system with the singular distribution of
the field. This generalized approach works for a large class of models. We use
this method in section \ref{sec:gyro} to describe the dynamics of gyroscopic
systems. Our approach is applied to the problem of collective-variable
Lagrangian description of gyroscopic systems; effective equation of motion are
constructed in section \ref{sec:Thiele}. For gyroscopic systems in
two--dimensional (2D) magnetism, we present in section \ref{sec:magnetism}
explicit results for different models. The connection between the gyroscopic
force and the singular one is discussed. We consider the possibility of using
the collective variable Lagrangian approach in the magnetic solitons dynamics.
We conclude in section \ref{sec:conclusion}.


\section{Energy and force--balance equations}
\label{sec:balance} %

We study the classical Lagrangian dynamics for the multicomponent field
$\vec{\Phi}(\vec{x},t)$ in the $(d+1)$ spacetime dimensions, which is
described by Euler--Lagrange equations:
\begin{equation} \label{eq:Euler-Lagrange} %
\frac{\delta L}{\delta \Phi_k} \equiv \frac{\partial \mathscr{L}}{\partial
\Phi_k} - \partial_\alpha \frac{\partial \mathscr{L}}{\partial
{\Phi_k}_{,\alpha}} =0,
\end{equation}
where $L=\int_{\mathcal D} {\rmd}\vec{x}\, \mathscr{L}
\bigl(\Phi_k;{\Phi_k}_{,\alpha}\bigr)$. Here and below Latin indices $k$, $l$
describe components of the field $\vec{\Phi}$, Latin indices
$i,j=\overline{1,d}$ numerate spatial coordinates $x_i$, and Greek indices
$\alpha,\beta=\overline{0,d}$ correspond to spacetime coordinates.

We start with the standard definition of the field momentum
\begin{equation} \label{eq:momentum} %
\vec{P} = -\int_{\mathcal{D}}{\rmd}\vec{x} \frac{\partial
\mathscr{L}}{\partial {\Phi_k}_{,0}} \vec{\nabla} {\Phi_k}.
\end{equation}
To describe the dynamics of the system as a whole on the basic of the momentum
\eqref{eq:momentum}, let us consider the energy--momentum tensor
\cite{LandauII}
\begin{equation} \label{eq:T} %
T_{\alpha\beta} = {\Phi_k}_{,\alpha}\frac{\partial \mathscr{L}}{\partial
{\Phi_k}_{,\beta}} - \mathscr{L} \delta_{\alpha\beta}.
\end{equation}
The flux of the energy--momentum tensor can be calculated in the standard way:
\begin{equation} \label{eq:div-T-general} %
\partial_\beta T_{\alpha\beta} = {\Phi_k}_{,\alpha,\beta}
\frac{\partial \mathscr{L}}{\partial {\Phi_k}_{,\beta}} +
{\Phi_k}_{,\alpha}\partial_\beta \frac{\partial \mathscr{L}}{\partial
{\Phi_k}_{,\beta}} - \partial_\alpha \mathscr{L}.
\end{equation}
Calculating the derivative $\partial_\alpha \mathscr{L}$ taking into account
\eqref{eq:Euler-Lagrange} and changing the order of the derivation as follows
${\Phi_k}_{,\alpha, \beta} = {\Phi_k}_{,\beta, \alpha}$, we obtain the
well--known equation \cite{LandauII}
\begin{equation} \label{eq:div-T-reg} %
\partial_\beta T_{\alpha\beta}=0.
\end{equation}
In the integral form equation \eqref{eq:div-T-reg} with $\alpha=0$ corresponds
to the work equation for the total energy $\mathcal{E}=\int_{\mathcal D}
{\rmd}\vec{x} T_{00}$,
\begin{equation} \label{eq:energy-reg} %
\frac{{\rmd}\mathcal{E}}{{\rmd}t} = -\oint_{\partial\mathcal{D}}\!\!
{\rmd}f_i\, \mathcal{S}_i,
\end{equation}
which means that the energy changes due to the flux through the boundary,
$\mathcal{S}_i = T_{0i}$. Space components of the integral equation following
from \eqref{eq:div-T-reg} give the Newtonian--like equation
\begin{equation} \label{eq:force-balance-reg} %
\fl \frac{{\rmd}\vec{P}}{{\rmd}t}=\vec{F}^{\mathrm{reg}}, \quad
F_i^{\mathrm{reg}} = -\oint_{\partial \mathcal{D}}\!\! {\rmd}f_j\;
\varPi_{ij}, \quad\varPi_{ij} = \mathscr{L}\delta_{ij} -
{\Phi_k}_{,i}\frac{\partial \mathscr{L}}{\partial {\Phi_k}_{,j}}.
\end{equation}
Let us remind that to derive relations
\eqref{eq:div-T-reg}--\eqref{eq:force-balance-reg}, it is necessary to suppose
that second derivatives of any field components commute,
$[\partial_\alpha,\partial_\beta]\Phi_k=0$. This is a standard assumption,
which works well for the smooth distribution of the field $\vec{\Phi}$.
However, for the system with topological defects this assumption can fail.

Probably, the most familiar kind of singularity is the \emph{phase
singularity} \cite{Berry91}, which can be found in different physical systems.
In a light wave, the phase singularity is known as an optical vortex
\cite{Desyatnikov05}; such a singular phenomenon gives birth to the singular
optics. One of the well--known example of the phase singularity in the
condense matter physics is the 2D quantum Hall systems \cite{Simon98}, where
the Chern--Simons approach is employed by making the singular gauge
transformation on the phase of the electron wavefunction. In the simplest case
of a single electron, this transformation can be written as $\psi\to
\phi\cdot\psi$, where $\phi=\rme^{-\imath\arg(z-z_0)}$, and $z\in\mathbb{C}$
is a point in the $xy$ plane. This leads to the additional Chern--Simons
magnetic field with the vector potential (or the Berry connection in
accordance to \cite{Girvin99}) $\vec{A} = \imath\vec{\nabla}\phi$, which has a
singularity at $z=z_0$ due to the multivalued function $\arg(z)$. The
corresponding magnetic induction does not vanish at $z_0$, namely
$\vec{B}=\vec{\nabla}\times\vec{\nabla}\phi = 2\pi\delta(z-z_0) \vec{e}_3$,
where $\delta(z)$ is 2D Dirac's delta function. The same situation takes place
for the Aharonov--Bohm effect, where the Berry phase $\phi$ can be interpreted
as an Aharonov--Bohm phase \cite{Girvin99}. The singular field distribution
appears for 2D solitons in magnets, where the in--plane angle of magnetization
is described by the multivalued $\arg(z)$--function, $\phi=\arg(z-z_0)$
\cite{Kosevich90}. The second derivatives of $\phi$ do not commute,
$\epsilon_{ij}\partial_i\partial_j\phi = 2\pi\delta(z-z_0)$, this is well
discussed by \citet{Papanicolaou91}. All above mentioned singularities are
connected with Dirac's monopole: the vector potential has a Dirac string along
some direction (in our case the string crosses an $xy$--plane at $z_0$), which
breaks the invariance of the system \cite{Haldane86}.

Let us calculate the energy--momentum dynamics equations, which allow the
field singularities. Simple calculations taking into account
\eqref{eq:Euler-Lagrange}, \eqref{eq:T} and \eqref{eq:div-T-general} lead to
\emph{the generalized expression} for the energy--momentum flux:
\begin{equation} \label{eq:div-T} %
\partial_\beta T_{\alpha\beta}=\frac{\partial
\mathscr{L}}{\partial {\Phi_k}_{,\beta}}\left( {\Phi_k}_{,\alpha,\beta} -
{\Phi_k}_{,\beta,\alpha}\right).
\end{equation}
In general case there exist a nonzero flux of the energy--momentum, so the
conservation laws in the system can vanish. A similar picture, when the
energy--momentum tensor cannot be presented in the covariant form, takes place
in the general relativity \cite{LandauII}. In the fluid dynamics such a
singularity is known for vortices \cite{Batchelor87}. Below we discuss several
examples in the condense matter physics, in particular, in the magnetism,
where such a singularity is connected to the gyroscopical dynamics of
topological excitations.

Using \eqref{eq:div-T} one can derive the work equation in the form
\begin{equation} \label{eq:energy} %
\frac{{\rmd}\mathcal{E}}{{\rmd}t} = -\oint_{\partial\mathcal{D}}\!\!
{\rmd}f_i\; \mathcal{S}_i
 +\int_{\mathcal{D}}\!\!{\rmd}\vec{x}\; \frac{\partial \mathscr{L}}{\partial
{\Phi_k}_{,i}} \left({\Phi_k}_{,i,0} - {\Phi_k} _{,0,i}\right).
\end{equation}
The energy changes not only due to the flux through the boundary as in
\eqref{eq:energy-reg}. The second term on the right--hand side (RHS) of the
work equation \eqref{eq:energy} describes the energy changes due to field
singularities.

The space components of the integral form of \eqref{eq:div-T} can be presented
in the Newtonian way, similar to \eqref{eq:force-balance-reg},
\begin{subequations} \label{eq:force-balance-sing}
\begin{equation} \label{eq:force-balance} %
\frac{{\rmd}\vec{P}}{{\rmd}t}=\vec{F}, \qquad \vec{F} =
\vec{F}^{\mathrm{reg}}+\vec{F}^{\mathrm{sing}}.
\end{equation}
The force has two contributions: one of them, $\vec{F}^{\mathrm{reg}}$ can be
expressed as the current of the stress tensor $\varPi_{ij}$, see
\eqref{eq:force-balance-reg}. An additional \emph{singular force}
\begin{equation} \label{eq:F-sing} %
F^{\mathrm{sing}}_i = \int_{\mathcal{D}}\!\!{\rmd}\vec{x}\; \frac{\partial
\mathscr{L}} {\partial
{\Phi_k}_{,\beta}}\left({\Phi_k}_{,\beta,i}-{\Phi_k}_{,i,\beta}\right)
\end{equation}
\end{subequations}
appears only if the field distribution has a singularity (when derivatives of
$\vec{\Phi}$ are not smooth in $\mathcal{D}$). Namely, this additional force
$\vec{F}^{\mathrm{sing}}$ is the main issue of our investigation.

If the field distribution $\vec{\Phi}(\vec{x},t)$ is calculated, then equation
\eqref{eq:force-balance} describes the effective equation of motion for the
system as a whole. Such approach is known to be applied to the dynamics of
regular fields, where it takes the form \eqref{eq:force-balance-reg}, see
\cite{Kosevich90,Baryakhtar93}. Existence of the force
$\vec{F}^{\mathrm{sing}}$ is caused by the additional flux through the region
of the field singularity.

Let us discuss the possible candidates who admits effects of
$\vec{F}^{\mathrm{sing}}$. An explicit form of this force \eqref{eq:F-sing}
shows that it is absent for one--dimensional (1D) systems, where
$[\partial_0,\partial_1]{\Phi_k}=0$. That is why it is possible to use the
standard force balance \eqref{eq:force-balance-reg} for the description of the
dynamics of 1D solitons \cite{Kosevich90}.

Apparently, the singular force $\vec{F}^{\mathrm{sing}}$ can appear in
systems, where the Lagrangians contain non--potential terms, because the
energy density should be finite. Such properties have gyroscopical systems.
Therefore, the generalized force--balance equation \eqref{eq:force-balance}
taking into account singular force $\vec{F}^{\mathrm{sing}}$ should be used
for the description of the gyroscopic dynamics for systems with singular
topological solitons. Note that usage of the standard force balance in the
form \eqref{eq:force-balance-reg} leads to the discrepancy in the definition
of the gyroscopic force between the soliton perturbation theory
\cite{Nikiforov83} and direct integration of the field equations
\cite{Thiele73,Slonczewski79,Ivanov95d}.


\section{Gyroscopic systems in the field theory}
\label{sec:gyro} %

Let us consider the field system, whose dynamics has only gyroscopic
properties. The Lagrangian of such simplest gyroscopic system has the form
\begin{equation} \label{eq:Lagrangian-gyro} %
\mathscr{L}\left(\Phi_k;{\Phi_k}_{,\alpha}\right) =
\mathscr{G}-\mathscr{H}\equiv A_k(\vec{\Phi}) {\Phi_k}_{,0} - \mathscr{H}.
\end{equation}
We suppose that the `Hamiltonian' $\mathscr{H}$ is a regular function of
$\vec{\Phi}$ and $\vec{\Phi}_{,i}$, and all peculiarities can appear only due
to the gyroscopic term $\mathscr{G}=A_k{\Phi_k}_{,0}$. Such a form of the
Lagrangian corresponds to the case of a system with regular gyroscopic matrix,
which was systematically studied in \cite{Schnitzer00}, using a
collective-variable theory for constrained Hamiltonian systems of a classical
mechanics.

The Euler--Lagrange equations for this system have the form
\begin{equation} \label{eq:Euler-Lagrange4gyro} %
\mathcal{G}_{kl}{\Phi_l}_{,0} = \frac{\delta H}{\delta \Phi_k}
\end{equation}
with the antisymmetric gyroscopic tensor $\mathcal{G}_{kl} = {\partial
A_l}/{\partial \Phi_k} - {\partial A_k}/{\partial \Phi_l}$.

Let us calculate integral Newtonian equations in the form
\eqref{eq:force-balance}. The field momentum for the system
\eqref{eq:Lagrangian-gyro} has a gyroscopical nature,
\begin{equation*} \label{eq:momentum4gyro}
\vec{P}^{(g)} = -\int_{\mathcal{D}}{\rmd}\vec{x} A_k\vec{\nabla}\Phi_k.
\end{equation*}
Let us start with a regular field distribution, when
${\rmd}\vec{P}^{(g)}/{\rmd}t = \vec{F}^{\mathrm{reg}}$, see
\eqref{eq:force-balance-reg}. Supposing that the field distribution is also
localized, one can write the Newtonian equation in the form of the
force-balance condition:
\begin{equation} \label{eq:force-balance-reg-1}
\fl %
\vec{F}^{(g)} + \vec{F}^{\mathrm{reg}(H)} = 0, \quad F_i^{\mathrm{reg}(H)} =
\oint_{\partial \mathcal{D}}{\rmd} f_j \left(\mathscr{H} \delta_{ij} -
{\Phi_k}_{,i}\frac{\partial \mathscr{H}}{\partial {\Phi_k}_{,j}} \right).
\end{equation}
Here, the quantity
\begin{equation} \label{eq:gyroforce-standard}
\vec{F}^{(g)} = -\frac{{\rmd}\vec{P}^{(g)}}{{\rmd}t}
\end{equation}
is an `internal' gyroscopic force, which acts together with external force
$\vec{F}^{\mathrm{reg}(H)}$ on the system. The gyroscopic force in this form
was introduced in \cite{Ivanov89} and used after that for the description of
regular field distributions, see for the review \cite{Baryakhtar93}.

The picture drastically changes if we consider singular field distributions.
Let us write the force-balance equation \eqref{eq:force-balance}, separating
the gyroscopic contribution:
\begin{subequations} \label{eq:force-balance-sing-gyro} %
\begin{eqnarray} \label{eq:force-balance-sing-1} %
\frac{{\rmd}\vec{P}^{(g)}}{{\rmd}t} &=& \vec{F}^{\mathrm{reg}(g)} +
\vec{F}^{\mathrm{reg}(H)} + \vec{F}^{\mathrm{sing}(g)},\\
\label{eq:gyroforce-reg} %
F_i^{\mathrm{reg}(g)} &=& -\oint_{\partial \mathcal{D}} {\rmd}f_i A_k {\Phi_k}_{,0},\\
\label{eq:gyroforce-sing} %
F^{\mathrm{sing}(g)}_i &=& \int_{\mathcal{D}} {\rmd}\vec{x} A_k
\left({\Phi_k}_{,0,i}-{\Phi_k}_{,i,0}\right).
\end{eqnarray}
\end{subequations}
To fashion the Newtonian equation \eqref{eq:force-balance-sing-1} as the
force--balance condition \eqref{eq:force-balance-reg-1}, we define the
gyroscopic force as follows:
\begin{equation} \label{eq:gyroforce} %
\vec{F}^{(g)} = -\frac{{\rmd}\vec{P}^{(g)}}{{\rmd}t} +
\vec{F}^{\mathrm{reg}(g)} + \vec{F}^{\mathrm{sing}(g)}.
\end{equation}
This definition of the gyroscopic force differs from the usual one
\eqref{eq:gyroforce-standard}. Note that using the gauge transformation $A_k
\to A_k - A_k^{\mathrm{ground}}$ it is possible to suppress an effect of
$\vec{F}^{\mathrm{reg}(g)}$. Nevertheless, the presence of the singular force
\eqref{eq:gyroforce-sing} breaks the simple relation
\eqref{eq:gyroforce-standard}. Moreover, we will see below in equation
\eqref{eq:link} that for magnetic systems the gyroscopic force and the
singular one have the same value, $\left|\vec{F}^{(g)}\right| =
\left|\vec{F}^{\mathrm{sing}}\right|$.

One can rewrite the complicated expression \eqref{eq:gyroforce} for the
gyroscopic force in the compact form
\begin{equation} \label{eq:F-gyro-OK} %
F_i^{(g)} = \int_{\mathcal{D}} {\rmd}\vec{x} \mathcal{G}_{kl}{\Phi_k}_{,0}
{\Phi_l}_{,i},
\end{equation}
which can be used for the description both localized topological solitons
(skyrmions) \cite{Thiele73,Ivanov89,Baryakhtar93} and nonlocalized vortices
\cite{Nikiforov83,Wysin94a,Mertens00}.

It is easy to generalize results for systems, whose dynamics admit both
kinetic and gyroscopic properties. Let us start with the Lagrangian system
\begin{equation*} \label{eq:Lagrangian} %
\mathscr{L} = \mathscr{G} + \mathscr{L}^{(0)},
\end{equation*}
where we separate the gyroscopic term $\mathscr{G}$ from the Lagrangian and
suppose that $\mathscr{L}^{(0)}$ has no singularities. The simple
generalization of the force--balance relation takes a form
\begin{equation} \label{eq:dP/dt}
\fl \frac{{\rmd}\vec{P}^{(0)}}{{\rmd}t} = \vec{F}^{(0)} + \vec{F}^{(g)},\qquad
F_i^{(0)} = - \oint_{\partial \mathcal{D}} {\rmd}f_j \left(\mathscr{L}^{(0)}
\delta_{ij} - {\Phi_k}_{,i}\frac{\partial \mathscr{L}^{(0)}}{\partial
{\Phi_k}_{,j}} \right).
\end{equation}
It is instructive to mention an analogy with an equation of motions of the
charged particle $m$ in the electromagnetic field $\vec{A}$ under the action
of the external force $\vec{F}$. Since the canonical momentum of the particle
is $\vec{P} = m\vec{v} + \vec{A}$, the Newtonian equation of motion takes a
form
\begin{equation} \label{eq:electrod}
\frac{{\rmd}\vec{P}}{{\rmd}t} = \vec{F}, \quad \mathrm{or} \quad
m\frac{{\rmd}\vec{v}}{{\rmd}t} = \vec{F} - \frac{{\rmd}\vec{A}}{{\rmd}t}.
\end{equation}
The last term $-\rmd \vec{A}/{\rmd}t$ can be interpreted as a Lorentz force,
the particular case of a gyroscopic force, in analogy with the relation
$\vec{F}^{(g)} = -{{\rmd}\vec{P}^{(g)}}/{{\rmd}t}$, see
\eqref{eq:gyroforce-standard}. Note that the sign `minus' always appears in
the gyroscopic force, because internal gyroscopical properties of the whole
system (in the example \eqref{eq:electrod} this system consists of the
particle and the electromagnetic field) are interpreted as an additional
force, which acts on a particle.

\section{Thiele approach and effective Lagrangian}
\label{sec:Thiele} %

Let us consider the collective-variable dynamics of the gyroscopic system with
the Lagrangian \eqref{eq:Lagrangian-gyro}. The collective-variable description
becomes important in the nonlinear field theories, when the field distribution
has a well-defined particle-like properties. If the system admits the
travelling wave solution (\emph{travelling wave ansatz}, TWA),
\begin{equation} \label{eq:tr-wave-ansatz}
\Phi_k^{\mathrm{TWA}}(\vec{x},t) = \Phi_k\left(\vec{x}- \vec{X}(t)\right),
\end{equation}
one can derive the gyroscopic force \eqref{eq:F-gyro-OK} in the form
\begin{subequations} \label{eq:F-gyro4Thiele}
\begin{equation} \label{eq:F-gyro4Thiele-1}
F_i^{(g)} = G_{ij}\dot{X}_j.
\end{equation}
Here the gyroscopic tensor
\begin{equation} \label{eq:G}
G_{ij} = \int_{\mathcal{D}} {\rmd}\vec{x}
\mathcal{G}_{kl}{\Phi_k}_{,i}{\Phi_l}_{,j}
\end{equation}
\end{subequations}
is an extension  of the gyrocoupling tensor, obtained by \citet{Thiele73}, to
general gyroscopic systems. Then, the force-balance condition takes the form
of Thiele-like equations, cf. \cite{Thiele73}
\begin{equation} \label{eq:Thiele-eq}
G_{ij}\dot{X}_j + F_i(\vec{X})=0, \qquad \vec{F} \equiv
\vec{F}^{\mathrm{reg}(H)} = -\frac{\partial H}{\partial\vec{X}},
\end{equation}
where $H=\int_{\mathcal{D}}{\rmd}\vec{x}\,\mathscr{H}$. Note that one can
derive Thiele--like equations from the effective Lagrangian:
\begin{equation} \label{eq:Thiele-L-eff}
\fl L^{\mathrm{eff}} = \case{1}{2} G_{ij}X_i\dot{X}_j - H,\qquad G_{ij}=
\int_{\mathcal{D}} {\rmd}\vec{x} \left(\frac{\partial A_l}{\partial \Phi_k} -
\frac{\partial A_k}{\partial \Phi_l}\right) {\Phi_k}_{,i}{\Phi_l}_{,j}.
\end{equation}

The generalization of Thiele-like equations \eqref{eq:Thiele-eq} in the spirit
of collective-variable theory can be made for the case when there is no exact
travelling wave solution. The basis of this theory is a \emph{generalized
travelling wave ansatz} \cite{Mertens97,Mertens00}
\begin{equation*}
\Phi_k(\vec{x},t) =
\Phi_k\left(\vec{x}-\vec{X}(t),\partial_0\vec{X}(t),\partial_0^2\vec{X}(t),
\dots,\partial_0^n\vec{X}(t)\right),
\end{equation*}
which leads to the $(n+1)^{\mathrm{th}}$ order equation of motion:
\begin{equation} \label{eq:gen-Thiele-eq}
\sum_{k=1}^{n+1} G_{ij}^k \partial_0^kX_j + F_i(\vec{X}) = 0.
\end{equation}
Note that in the Thiele approximation $n=0$ and $G_{ij}^1=G_{ij}$.

Another kind of a generalization appears when internal degrees of freedom
become important. For example, in the Rice approach \cite{Rice83} for the 1D
Klein--Gordon model the kink width becomes a collective variable as well as
its position. Generalization for 2D solitons and vortices has been done
recently in \cite{Zagorodny04,Sheka05a}. That is why we will discuss here the
possibility of deriving the effective Lagrangian of the system directly by
integrating the microscopic Lagrangian \eqref{eq:Lagrangian-gyro} with the
travelling wave ansatz \eqref{eq:tr-wave-ansatz}.

Let us define the effective Lagrangian of the gyroscopic system:
\begin{equation*} \label{eq:L-eff-2}
L^{\mathrm{eff}} = \int_{\mathcal{D}}{\rmd}\vec{x}\,
\mathscr{L}\bigl(\Phi_k^{\mathrm{TWA}};
{\Phi_k}_{,\alpha}^{\mathrm{TWA}}\bigr).
\end{equation*}
It is easy to see that the effective momentum coincides with the standard
field momentum, calculated with the travelling wave ansatz, $\partial
L^{\mathrm{eff}}/\partial \dot{\vec{X}}=\vec{P}$. In the same way one can
calculate the effective force $\partial L^{\mathrm{eff}}/\partial {\vec{X}}$,
which is equal to the regular force $\vec{F}^{\mathrm{reg}}$. Thus effective
Euler--Lagrange equations have the form of the singular force-balance
condition \eqref{eq:force-balance-sing}, which can be presented as follows:
\begin{equation} \label{eq:L-eff-EoM-1}
\fl \frac{{\rmd}}{{\rmd} t} \frac{\partial L^{\mathrm{eff}}}{\partial
\dot{\vec{X}}} -\frac{\partial L^{\mathrm{eff}}}{\partial \vec{X}} =
\vec{F}^{\mathrm{sing}},\qquad F^{\mathrm{sing}}_i = \dot{X}_j
\int_{\mathcal{D}}{\rmd}\vec{x}\, A_k
\left({\Phi_k}_{,i,j}-{\Phi_k}_{,j,i}\right).
\end{equation}
The standard effective Lagrangian description is adequate \emph{only} when the
singular force is absent.

Let us consider the situation when we should obviate difficulties with the
singular force. The gauge transformation $A_k \to A_k+\partial
f(\vec{\Phi})/\partial \Phi_k$ changes the gyroscopic tensor by the value
\begin{equation*} \label{eq:G^gauge}
\mathcal{G}_{kl}^{\mathrm{gauge}} = \frac{\partial^2 f}{\partial
\Phi_k\partial \Phi_l} - \frac{\partial^2 f}{\partial \Phi_k\partial \Phi_l}.
\end{equation*}
If the function $f(\vec{\Phi})$ is smooth enough and the second derivatives
commute, the gauge transformation does not change equations of motions
\eqref{eq:Euler-Lagrange4gyro}. Nevertheless, there could appear uncertainty
in the canonic momentum definition. Under the gauge transformation, the
momentum changes by the value
\begin{equation*}
\vec{P}^{\mathrm{gauge}} = -\int_{\mathcal{D}}{\rmd}\vec{x} \frac{\partial
f}{\partial \Phi_k} \vec{\nabla}\Phi_k,
\end{equation*}
which is not well--define for the singular field distributions
\cite{Haldane86,Volovik03}. If $A_k(\vec{\Phi}(\vec{x}))$ takes the value
$A_k^{\mathrm{sing}} = A_k(\vec{\Phi}(\vec{x}_0))$ in a singular point
$\vec{x}_0$ of the field $\vec{\Phi}$, then after the gauge transformation
$A_k\to A_k-A_k^{\mathrm{sing}}$ the Lagrangian \eqref{eq:Lagrangian-gyro}
will have no singularity. Thus, the following effective Lagrangian approach is
valid,
\begin{equation} \label{eq:L&EoM-reg}
\fl L^{\mathrm{eff}} = \int_{\mathcal{D}}{\rmd}\vec{x} \left[
\left(A_k-A_k^{\mathrm{sing}}\right){\Phi_{k,0}^{\mathrm{TWA}}} - \mathscr{H}
\right], \qquad \frac{{\rmd}}{{\rmd} t} \frac{\partial
L^{\mathrm{eff}}}{\partial \dot{\vec{X}}} -\frac{\partial
L^{\mathrm{eff}}}{\partial \vec{X}} = 0.
\end{equation}
Such an approach can be generalized for the case when the field $\vec{\Phi}$
has several singular points $\vec{x}_n$, but with the same behaviour,
$A_k^{\mathrm{sing}} = A_k(\vec{\Phi}(\vec{x}_n))$. To illustrate this
effective Lagrangian method \eqref{eq:L&EoM-reg} we construct below an
effective Lagrangian for the magnetic vortex dynamics, see
\eqref{eq:L-eff-with-p}.


\section{Application to the 2D magnetism}
\label{sec:magnetism} %

In this section, we apply our results to the dynamical properties of 2D
topological defects (solitons and vortices) in magnetic systems. In the
continuum limit, the dynamics of the broad class of Heisenberg magnets can be
described in terms of the unit order parameter vector $\vec{n} =
\left(\sin\theta\cos{\phi}, \sin\theta\sin{\phi}, \cos\theta\right)$; for the
classical ferromagnet $\vec{n}$ is the normalized magnetization, for the
antiferromagnet $\vec{n}$ is the normalized sublattice magnetization vector.
Thus, the magnet can be described by the two--component field $\vec{\Phi} =
\left( \theta, {\phi}\right)$.

Let us start with the case of the ferromagnet, whose dynamics is described by
Landau--Lifshitz equations \cite{Kosevich90}. Using $\pi\equiv\cos\theta$ as a
canonical momentum for the azimuthal angle $\phi$, dynamical equations take
the form
\begin{equation} \label{eq:LL}
\dot{\phi}=\frac{\delta H}{\delta\pi}, \qquad \dot{\pi} = - \frac{\delta
H}{\delta\phi}.
\end{equation}
Note that in spite of the fact that $\pi$ and $\phi$ do have a form of a
canonic pair, these variables are not well defined \cite{Volovik03}: the
azimuthal angle $\phi$ is ill--defined when $\theta=0,\pi$.

Topological properties of solutions are determined by the mapping of the
$xy$--plane to the $S^2$--sphere of the order parameter space. This mapping is
described by the homotopic group $\pi_2(S^2) = \mathbb{Z}$, which is
characterized by the topological invariant (Pontryagin index):
\begin{equation} \label{eq:Pontryagin}
Q=\frac 1{4\pi }\int\!\!{\rmd}^2x\, \mathcal{Q}, \qquad
\mathcal{Q}=\epsilon_{ij}\pi_{,i}\phi_{,j}.
\end{equation}
The Pontryagin index takes integer values, $Q\in\mathbb{Z}$, being an integral
of motion.

In the Lagrangian approach one can derive Landau--Lifshitz equations
\eqref{eq:LL} from the functional
\begin{equation} \label{eq:Lagrangian4FM}
L = - \int {\rmd}^2x \bigl(C-\cos\theta\bigr)\partial_{0}\phi - H,
\end{equation}
where $C$ is an arbitrary constant \cite{Nikiforov83}. Usually, one chooses
$C=1$ in order to neglect the contribution of the ground state (which
corresponds to $\theta=0$ for easy--axis magnets) \cite{Kosevich90}. Then, the
standard definition of the ferromagnet momentum integral reads
\begin{equation} \label{eq:momentum-FM}
\vec{P} = \int {\rmd}^2x (1-\cos\theta)\vec{\nabla}\phi.
\end{equation}
Note that namely this definition of the momentum is the origin of the
long--time discussion in the literature
\cite{Haldane86,Volovik87,Papanicolaou91,Wysin94a,Mertens00}. The main
criticism of the momentum definition \eqref{eq:momentum-FM} is connected with
its conservation. It was shown by \citet{Haldane86} that the momentum
\eqref{eq:momentum-FM} cannot be conserved for a singular field distribution.
The origin is in the singularity of the Lagrangian at some point. In order to
visualize the singularity, let us rewrite the Lagrangian
\eqref{eq:Lagrangian4FM} using the dimension quantity $\vec{M}=M\vec{n}$
without constrain $\vec{M}^2=\mathrm{const}$,
\begin{equation} \label{eq:A-monopole}
\mathscr{L} = \vec{A} \cdot \partial_{0}\vec{M} - \mathscr{H}, \qquad \vec{A}
= \frac{\left[\vec{n}_0 \times \vec{M}\right]}{M(M+\vec{n}_0\cdot\vec{M})}.
\end{equation}
Here, $\vec{A}$ is the vector potential of effective magnetic field
\cite{Ivanov04a}. One can see that $\vec{A}$ has a singularity along the line
$\vec{n}_0\cdot\vec{M} = -M$. It is easy to calculate the magnetic induction
of the `magnetic field', $\vec{B}=\vec{\nabla}_{\!\!\vec{M}}\times \vec{A} =
-\vec{M}/M^3$, which coincides with a magnetic induction of a Dirac magnetic
monopole. Thus, the vector potential $\vec{A}$ has a Dirac string along the
direction $\vec{n}_0$, which breaks the rotation invariance of the model
\cite{Haldane86}.

Since the Lagrangian \eqref{eq:Lagrangian4FM} has a singularity, the standard
momentum \eqref{eq:momentum-FM} is not well defined
\cite{Haldane86,Volovik87}, moreover it is not conserved. That is why
\citet{Papanicolaou91} proposed another definition of the momentum, which is
connected only with the topological properties of the ferromagnet
\eqref{eq:Pontryagin},
\begin{equation} \label{eq:momentum-Nikos}
P_i^{\mathrm{PT}} = \epsilon_{ij}\int {\rmd}^2x\,  x_j \mathcal{Q}.
\end{equation}
The momentum \eqref{eq:momentum-Nikos} is an analogue of a fluid impulse,
which is defined as a linear moment of a local vorticity and used for the
description of the fluid vortex dynamics \cite{Batchelor87}.

The Poisson bracket relation for the momentum \eqref{eq:momentum-Nikos} takes
the nonzero value, $\bigl\{P_1^{\mathrm{PT}},P_2^{\mathrm{PT}}\bigr\}=4\pi Q$
as well as for the standard momentum \eqref{eq:momentum-FM},
$\bigl\{P_1,P_2\bigr\}=4\pi Q$. An advantage of the momentum definition is its
conservation for the finite energy field distribution \cite{Papanicolaou91}.

The momentum $\vec{P}^{\mathrm{PT}}$ is claimed in \cite{Papanicolaou91} to be
a generator of space translations. However, as it was shown in
\cite{Wysin94a}, the Poisson bracket between $\vec{P}^{\mathrm{PT}}$ and any
smooth functional $F[\phi(\vec{x}),\pi(\vec{x})]$ takes the form
\begin{equation} \label{eq:Poisson4Nikos}
\fl %
\bigl\{P_i^{\mathrm{PT}},F\bigr\} = -\int\! {\rmd}^2x \left(\phi_{,i}
\frac{\delta F}{\delta \phi} + \pi_{,i} \frac{\delta F}{\delta \pi}\right) +
\epsilon_{ij}\epsilon_{kl} \int\! {\rmd}^2x\, x_j\phi_{,k,l}\frac{\delta
F}{\delta \phi}.
\end{equation}
Thus, $\vec{P}^{\mathrm{PT}}$ defines a true momentum functional only if the
last term in \eqref{eq:Poisson4Nikos} vanishes. It seems to vanish due to the
antisymmetric tensor $\epsilon_{kl}$ is constrained with the symmetric
$\phi_{,k,l}$. However, it is valid only for the regular field distribution.
As an example let us consider the following singular distribution of the
field, which corresponds to the simplest 2D topological defect
\begin{equation} \label{eq:2D-soliton} %
\theta=\theta(|z-z_0|),\qquad {\phi}=Q\cdot \mathrm{arg\ }(z-z_0),
\end{equation}
where $z=x+\imath\,y$, and $z_0\in\mathbb{C}$ is the position of the centre of
the defect. The singular properties appear for the field variable ${\phi}$:
the second derivatives do not commute, $\epsilon_{kl}{\phi}_{,k,l}=2\pi
Q\delta(z-z_0)$. The last term in \eqref{eq:Poisson4Nikos} finally reads $2\pi
Q \epsilon_{ij}x_j \delta F/\delta\phi\Bigr|_{z=z_0}$. One can see that in
general the momentum $\vec{P}^{\mathrm{PT}}$ can not be the translation
generator.

Also note that the momentum $\vec{P}^{\mathrm{PT}}$ does not describe an
individual soliton dynamics. Using the algebra for the momentum
$\vec{P}^{\mathrm{PT}}$, it was shown in \cite{Papanicolaou91} that a single
topological structure cannot move in the absence of an external field: the
soliton with $Q\neq0$ is always pinned at some point in $xy$--plane; it is
possible to move the set of solitons \cite{Papanicolaou95}. This does not
prevent the rotation motion of the soliton; however, the centre of the soliton
orbit is fixed \cite{Papanicolaou91}. In this context, let us mention an
analogy with the cyclotron motion of the electron: the electron coordinate
changes when it moves along the cyclotron orbit, its standard momentum also
changes, while their combination, the guiding centre position, is conserved.
Namely, this guiding centre coordinate corresponds to the momentum
$\vec{P}^{\mathrm{PT}}$, see \cite{Papanicolaou91}.

One can see that the momentum $\vec{P}^{\mathrm{PT}}$ cannot provide an
information about an instant soliton position, while the standard definition
of the momentum gives a possibility of describing a single soliton motion,
because it determines the gyroscopic force \eqref{eq:gyroforce} and depends on
the instant soliton position $\vec{X}(t)$, see \eqref{eq:F-gyro4Thiele-1}. The
possibility of a single soliton motion was predicted in \cite{Sheka01} for the
easy--axis ferromagnet: it results from the complicated internal structure of
the soliton. This motion was observed in simulations recently \cite{Sheka06}
by exciting a certain magnon mode, localized on the soliton.

Therefore, we go back to the standard definition of the ferromagnet momentum
\eqref{eq:momentum-FM}. Let us start with the localized distribution of the
magnetization field, which corresponds to the magnetic skyrmion. We discuss
here the problem of the conservation of the momentum \eqref{eq:momentum-FM},
when an external force is absent, $\vec{F}^{\mathrm{reg}}=0$. Using the
force--balance equations \eqref{eq:force-balance-sing-gyro}, one can derive
\begin{equation} \label{eq:force-balance-magn}
\frac{{\rmd}P_i^{(g)}}{{\rmd}t} = F_i^{\mathrm{sing}} = \int_{\mathcal{D}}
{\rmd}^2x (\cos\theta-1)\left(\phi_{,0,i}-\phi_{,i,0}\right).
\end{equation}
The total momentum is conserved only if $\vec{F}^{\mathrm{sing}}=0$. However,
the singular force does not vanish even for the simplest case of the
Thiele--like motion of the soliton, which has a structure
\eqref{eq:2D-soliton}
\begin{equation*} \label{eq:F-sing4FM}
F_i^{\mathrm{sing}} = \epsilon_{ij}\varGamma \dot{X}_j, \qquad \varGamma =
-4\pi Q,
\end{equation*}
where we suppose that in the centre of the soliton $\cos\theta=-1$. Let us
calculate the gyroscopic force $\vec{F}^{(g)}$, which acts on the soliton from
the media. Using \eqref{eq:F-gyro4Thiele}, one can present the gyroscopic
force in the form $F_i^{(g)}=\epsilon_{ij}G \dot{X}_j$, where the gyroscopic
constant $G=4\pi Q$. Thus, the gyroscopic force is caused by the field
singularity, $G=-\Gamma$.

\begin{table}
\caption{\label{table:gyro} %
Gyroscopic coefficients for different magnets: %
1) easy axis (EA) and isotropic ferromagnet (FM) with spin $S$ and
lattice constant $a$ \cite{Kosevich90}; %
2) easy--plane (EP) FM in the perpendicular magnetic field $h=H/H_a$,
parameter $p\equiv\cos\theta(0) =\pm1$ describes the polarity of the vortex
\cite{Nikiforov83}; 3) EP antiferromagnet (AFM) in the perpendicular field $H$
\cite{Ivanov94a}.
} %
\begin{indented}
\item[]
\begin{tabular}{@{}llll}
\br Type of magnet & Type of defect & $A(\theta)$ & $G$ \\
\mr
1) EA FM & Solitons & $\hslash Sa^{-2}(1-\cos\theta)$ & $-4\pi Q\hslash S/a^{2}$   \\
2) EP FM & Vortex & $\hslash Sa^{-2}\cdot (h-\cos\theta)$ & $-2\pi Q(p-h)\hslash S/a^{2}$ \\
3) EP AFM & Vortex & $-(gH/c^2)\cdot\cos^2\theta$ & $2\pi Q\cdot (gH/c^2)$                 \\
\br
\end{tabular}
\end{indented}
\end{table}

We have considered the case of magnetic skyrmions. It is possible to
generalize results for different 2D topological defects. Let us consider the
case of uniaxial 2D magnets, whose gyroscopic properties can be described by
the following gyroscopic term in the Lagrangian: $\mathscr{G} =
A(\theta)\partial_{0}\phi$. The form of the function $A(\theta)$ depends on
the magnet type, see table \ref{table:gyro}, where we mention only models,
which admit gyroscopic effects.

The simplest static nonlinear excitation in 2D magnets is the soliton for the
isotropic and easy--axis magnets and the vortex for the easy--plane magnet.
The structure of these different topological defects can be described by the
field distributions \eqref{eq:2D-soliton}. For standard models of the
Heisenberg magnet all spatial derivatives ${\partial
\mathscr{L}}/{\partial{\phi}_{,i}}$ vanish in the singularity point $z_0$,
which is the centre of the defect; it agrees with arguments that the energy
density should be finite. Therefore only the time derivative can influence the
picture. The singular force takes the form, cf. \eqref{eq:gyroforce-sing},
\begin{equation} \label{eq:F-sing4magn} %
F^{\mathrm{sing}}_i = \int_{\mathcal{D}}{\rmd}^2x A(\theta)
\left({{\phi}}_{,0,i}- {{\phi}}_{,i,0}\right).
\end{equation}
For the steady--state Thiele--like motion \eqref{eq:tr-wave-ansatz} this
singular force
\begin{equation*} \label{eq:F-sing-concrete} %
F_i^{\mathrm{sing}} = \epsilon_{ij} \varGamma \dot{X}_j, \qquad \varGamma=2\pi
Q A(z\to z_0).
\end{equation*}
One can see that $\vec{F}^{\mathrm{sing}}$ has a gyroscopical behaviour. The
gyroscopic force \eqref{eq:F-gyro4Thiele-1} is determined by the gyroscopic
tensor \eqref{eq:G}
\begin{equation*}  \label{eq:F-gyro-concrete} %
F_i^{(g)} = \epsilon_{ij} G \dot{X}_j, \qquad G = 2\pi Q \bigl[A(z\to\infty) -
A(z\to z_0)\bigr].
\end{equation*}
On the first view the gyroscopic constant is determined by the topological
properties only: $G=-4\pi Q\cdot{\hslash S}/{a^2}$ for the soliton in the
isotropic magnet and $G=-2\pi pQ\cdot{\hslash S}/{a^2}$ for the vortex in the
easy--plane magnet, see table \ref{table:gyro}. However, if we switch on an
external magnetic field, the gyroscopic constant $G$ becomes a smooth function
of the magnetic field, namely $G\propto (p-h)$ in the case of the cone--state
ferromagnet and $G\propto H$ in the case of the antiferromagnet, see table
\ref{table:gyro}. In general, the gyroscopic force is determined not only by
the field distribution in the origin of the topological singularity, but also
by the field distribution far from their. However one can normalize the
quantity $A$ by the ground value, $A\to A-A(z\to\infty)$. Finally, the
gyroscopic force reads
\begin{equation} \label{eq:link} %
\vec{F}^{(g)} = -\vec{F}^{\mathrm{sing}}.
\end{equation}
This is an important relation between the gyroscopic force and the singular
force, which assists to avoid the discrepancy between different approaches in
the study of the gyroscopic properties of 2D magnetic solitons and vortices
\cite{Nikiforov83,Papanicolaou91,Wysin94a,Mertens00}.

Let us discuss the possibility of using the collective--variable Lagrangian
approach in the magnetic vortex dynamics. Usually, the Lagrangian of the
ferromagnet is taken in the form \eqref{eq:Lagrangian4FM} with
$C=\cos\theta(\infty)$. For the easy--plane magnet $C=0$, and $L=G-H$ with the
gyroscopic term
\begin{equation} \label{eq:G4vortex}
G = \int{\rmd}^2x \cos\theta\partial_{0}\phi.
\end{equation}
The field momentum $\vec{P} = - \int{\rmd}^2x \cos\theta\vec{\nabla}\phi$ is
commonly used in the magnetic vortex dynamics \cite{Mertens00}. Let us
consider the vortex in the circular 2D magnet of the radius $L$. It is well
known \cite{Mertens00} that the vortex in such a system rotates about the
system centre due to the competition between the gyroscopic force and the
image force, which imitates the interaction with a boundary. To describe the
vortex motion in the Thiele approach, one needs to elaborate the model with
the travelling wave ansatz \eqref{eq:tr-wave-ansatz}. Simple calculations show
that the gyroscopic term \eqref{eq:G4vortex} disappears after the integration
\cite{Sheka05a}\footnote{See equation (B7) in \cite{Sheka05a}, where the
gyroscopic term $\mathcal{G}_2$ corresponds to $G$ in our equation
\eqref{eq:G4vortex}.}, and the effective Lagrangian \emph{does not} contain
the gyroscopic term, $L=-H$. Thus, Euler--Lagrangian equations cannot provide
the well--known vortex rotation. The reason is the influence of the singular
force: the Euler--Lagrangian equation for the effective Lagrangian contains an
extra term $\vec{F}^{\mathrm{sing}}$, see \eqref{eq:L-eff-EoM-1}. We have
derived this force above, it is opposite to the gyroscopic force, see
\eqref{eq:link}.

In some cases it is possible to suppress the singularity and to construct the
Lagrangian directly by integrating the field Lagrangian. We can do it formally
as described in \eqref{eq:L&EoM-reg}; but in order to visualize the
singularity, let us consider the gauge transformation `$\cos\theta \to
\cos\theta + \mathrm{const}$' in the model \eqref{eq:G4vortex}. Under this
transformation, the Lagrangian changes by the value
\begin{equation} \label{eq:L-gauge}
L^{\mathrm{gauge}} =  \mathrm{const} \int_{\mathcal{D}}\!\! {\rmd}^2x\,
\partial_{0}\phi,
\end{equation}
which should not influence the Euler--Lagrange equations. However, the
function $\phi$ is not differentiable, and this integral does not vanish: one
can derive \eqref{eq:L-gauge} using the travelling wave ansatz
\eqref{eq:tr-wave-ansatz}. After integrating, we obtain the gauge term in the
form $L^{\mathrm{gauge}}=\mathrm{const}\, \pi Q \epsilon_{ij}X_i\dot{X}_j$,
cf. \cite{Sheka05a}. Thus, using the singular gauge transformation it is
possible to suppress the singular force effect. In the case of the magnetic
vortex with polarity $p$ (see the table \ref{table:gyro}), one can choose the
regular effective Lagrangian
\begin{equation} \label{eq:L-eff-with-p}
L^{\mathrm{eff}} =
\int_{\mathcal{D}}{\rmd}^2x\Bigl\{\bigl[\cos\theta^{\mathrm{TWA}}
-p\bigr]\partial_{0}\phi^{\mathrm{TWA}} - \mathscr{H} \Bigr\}.
\end{equation}
More general, the regularized gyroscopic term for the 2D magnetic system can
be presented in the form, cf. \eqref{eq:L&EoM-reg}
\begin{equation*} \label{eq:G-eff-magn}
\mathscr{G}=
\Bigl[A\bigl(\theta^{\mathrm{TWA}}(z)\bigr)-A\bigl(\theta^{\mathrm{TWA}}(z_0)\bigr)
\Bigr] \partial_{0}\phi^{\mathrm{TWA}}.
\end{equation*}
We should note that such simple picture works well when all singularities have
the same behaviour, i.e. when $\theta$--field takes the same value at all
singular points. This can fail, e.g., for the system of two opposite polarized
vortices \cite{Nikiforov83}. In this case it is necessary to take into account
singular force effects in the form of \eqref{eq:L-eff-EoM-1}.


\section{Conclusion}
\label{sec:conclusion} %

In conclusion, we have constructed the equation of motion, involving the force
and the momentum, which is suitable for the description of singular objects
like topological defects. This equation is a consequence of a more general
problem of a noncovariance of the energy--momentum tensor \eqref{eq:div-T}.
One of the well--known example of such a problem is the general relativity,
when the energy--momentum tensor for the gravitational field cannot be
presented in the covariant form \cite{LandauII}. Another example is the fluid
dynamics, where the Lagrangian principle is violated in the effective theory
\cite{Batchelor87}. In the condense matter physics, in particular, in the
magnetism, this nonlocality leads to the problem in the momentum definition.
The reason of such paradoxes comes from the fact that description of the
many--body system in terms of few fields is always approximate, see the
discussion in \cite[chapter 6]{Volovik03}. In the paper we did not describe
the microscopic theory; however, we have shown how and when it is possible to
resolve the problem using the energy--momentum tensor in the framework of the
generalized expression \eqref{eq:div-T} for the energy--momentum flux. We
prove that in addition to the regular force there appears the singular one
\eqref{eq:F-sing}. Effects of the such a force are important for gyroscopical
systems. We considered the gyroscopic dynamics of the classical field with
topological defects and established the relation \eqref{eq:gyroforce} between
the gyroscopic force, singular force and the time derivative of a standard
field momentum. We have applied our approach to describe the gyroscopic
properties of 2D topological defects (soliton and vortices) in 2D magnets and
presented explicit results for different models. An important relation
\eqref{eq:link} is established between the gyroscopic force and the singular
one: it shows that the gyroscopic properties are caused by the field
singularity, which avoids contradictions between different approaches
\cite{Nikiforov83,Papanicolaou91,Wysin94a,Mertens00}. Using the singular force
effects we also discuss the possibility of effective Lagrangian description,
using collective coordinate' approach with an application for magnetic soliton
and vortex dynamics in 2D magnets.

\ack

The question addressed in the paper arose in a discussion with Boris Ivanov. I
thank Franz Mertens and Dmitry Sheka for the discussion both the physics and
the text. This work was supported by the Alexander von Humboldt Foundation.

{\scriptsize


}
\end{document}